\documentclass[graybox]{svmult}
\usepackage{mathptmx}       % selects Times Roman as basic font
\usepackage{helvet}         % selects Helvetica as sans-serif font
\usepackage{courier}        % selects Courier as typewriter font
\usepackage{type1cm}        % activate if the above 3 fonts are
\usepackage{makeidx}         % allows index generation
\usepackage{graphicx}        % standard LaTeX graphics tool
\usepackage{multicol}        % used for the two-column index
\usepackage[bottom]{footmisc}% places footnotes at page bottom

\usepackage{epsfig}
\usepackage{amsmath}
\usepackage{amsfonts}
\usepackage{amssymb}
\usepackage{color}
\usepackage{epic}
\usepackage{tikz}
\usetikzlibrary{calc}
\usetikzlibrary{positioning,shadows.blur}
\usepackage{pifont}

\usepackage{pgfplots}
\usepackage{siunitx}
\usepackage[export]{adjustbox}
\usepackage{cite}
\usepackage{comment}
\makeindex             % used for the subject index
\newcommand{\isdef}{\mathrel{\mathrel{\mathop:}=}}

\definecolor{SteelBlue}{rgb}{0.27, 0.51, 0.71}
\definecolor{LightSteelBlue}{rgb}{0.6902,    0.7686 ,   0.8706}
\begin{document}
%%%%%%%%%%%%%%%%%%%%%%%%%%%%%%%%%%%%%%%%%%%%%%%%%%%%%%%%%%%%%%%%%%%%%%%%%%%%%%%%%%%%%%%%%
\title*{Simulation of morphogen and tissue dynamics}
\author{M.\ D.\ Peters, L.\ D.\ Wittwer, A.\ Stopka, D.\ Barac, C.\ Lang, D.\ Iber}
\institute{Peters, Wittwer, Stopka, Barac, Lang, Iber \at Department of Biosystems Science and Engineering, \email{dagmar.iber@bsse.ethz.ch}}
\maketitle
%%%%%%%%%%%%%%%%%%%%%%%%%%%%%%%%%%%%%%%%%%%%%%%%%%%%%%%%%%%%%%%%%%%%%%%%%%%%%%%%%%%%%%%%%
\abstract{Morphogenesis, the process by which an adult organism emerges from 
a single cell, has fascinated humans for a long time. Modelling this process can 
provide novel insights into development and the principles that orchestrate \
the developmental processes. This chapter focusses on the mathematical description 
and numerical simulation of developmental processes. 
In particular, we discuss the mathematical representation of morphogen 
and tissue dynamics on static and growing domains, 
as well as the corresponding tissue mechanics. 
In addition, we give an overview of numerical methods that are routinely used 
to solve the resulting systems of partial differential equations.
These include the finite element method and the Lattice Boltzmann method 
for the discretisation as well as 
the arbitrary Lagrangian-Eulerian method and the Diffuse-Domain method 
to numerically treat deforming domains.
}
%%%%%%%%%%%%%%%%%%%%%%%%%%%%%%%%%%%%%%%%%%%%%%%%%%%%%%%%%%%%%%%%%%%%%%%%%%%%%%%%%%%%%%%%%
\section{Introduction}
During morphogenesis, the coordination of the processes that control size,
shape, and pattern is essential to achieve stereotypic outcomes and comprehensive 
functionality of the developing organism.  There are two main components contributing 
to the precisely orchestrated process of morphogenesis: 
morphogen dynamics and tissue dynamics. 
While signalling networks control cellular behaviour, 
such as proliferation and differentiation, tissue dynamics in turn modulate diffusion, 
advection and dilution, and affect the position of morphogen sources and sinks. 
Due to this interconnection, the regulation of those processes is very complex. 
Although a large amount of experimental data is available today, many of  the underlying 
regulatory mechanisms are still unknown. In recent years, 
cross-validation of numerical simulations with experimental data has emerged 
as a powerful method to achieve  an integrative understanding of the complex 
feedback structures underlying morphogenesis, 
see e.g.\ \cite{iber_inferring_2011,iber_image-based_2015,gomez2017image, mogilner_modeling_2011,sbalzarini_modeling_2013}. 

Simulating morphogenesis is challenging because of the multi-scale nature of the process.
The smallest regulatory agents, proteins, measure only a few nanometers in diameter, 
while animal cell diameters are typically at least a 1000-fold larger 
cp.\ \cite{alberts2014molecular,VanDenHurk2005}, and developing organs start as a small 
collection of cells, but rapidly develop into structures comprised of ten thousands of 
cells, cf.\ \cite{Michael:WSI13}. 
A similar multiscale nature also applies to the time scale. 
The basic patterning processes during morphogenesis typically proceed within days. 
Gestation itself may take days, weeks, or months - in some cases even years,
see \cite{Michael:Ric10}.
Intracellular signalling cascades, on the other hand, may be triggered within seconds, 
and mechanical equilibrium in tissues can be regained in less 
than a minute after a perturbation, see e.g.\ \cite{Michael:LMG16}. 
The speed of protein turn-over, see \cite{Michael:EGNI+11}, 
and of transport processes, see e.g.\ \cite{muller2013morphogen}, falls in between. 
Together, this results in the multiscale nature of the problem. 

Given the multiscale nature of morphogenesis, 
combining signalling dynamics with tissue mechanics in the same computational 
framework is a challenging task. Where justified, models of morphogenesis approximate 
tissue as a continuous domain. In this case, patterning dynamics can be described 
by reaction-advection-diffusion models. Experiments have shown that a tissue can be 
well approximated by a viscous fluid over long time scales, i.e.\ several minutes to hours, 
and by an elastic material over short time scales, i.e.\ seconds to minutes, 
cf.\ \cite{Michael:FFSS98}. Accordingly, tissue dynamics can be included by using 
the Navier-Stokes equation and/or continuum mechanics. 
In addition, cell-based simulation frameworks of varying resolution have been 
developed to incorporate the behaviour of single cells. 
These models can be coupled with continuum descriptions where appropriate.

In this review, we provide an overview of approaches to describe,
couple and solve dynamical models that represent tissue mechanics and signalling networks.
Section~\ref{sec:morphTissue} deals with the mathematical 
representation of morphogen dynamics, tissue growth and tissue mechanics. 
Section~\ref{sec:CellBased} covers cell-based simulation frameworks. 
Finally, Section~\ref{sec:NumApp} presents common numerical approaches to solve the respective models.
%%%%%%%%%%%%%%%%%%%%%%%%%%%%%%%%%%%%%%%%%%%%%%%%%%%%%%%%%%%%%%%%%%%%%%%%%%%%%%%%%%%%%%%%%
\section{Mathematical representation of morphogen and tissue dynamics}\label{sec:morphTissue}
\subsection{Morphogen dynamics}
A fundamental question in biology is that of self-organisation, 
or how the symmetry in a seemingly homogeneous system 
can be broken to give rise to stereotypical patterning and form. 
In 1952, Alan Turing first introduced the concept of a \textit{morphogen} 
in his seminal paper ``The Chemical Basis of Morphogenesis'', 
cf.\ \cite{Lucas:Turing1990-li}, in the context of self-organisation and patterning. 
He hypothesized that a system of chemical substances 
``reacting together and diffusing through a tissue'' was sufficient 
to explain the main phenomena of morphogenesis.

Morphogens can be transported from their source to target tissue in different ways. 
Transport mechanisms can be roughly divided into two categories: extracellular 
diffusion-based mechanisms and cell-based mechanisms \cite{muller2013morphogen}. 
In the first case, morphogens diffuse throughout the extracellular domain. 
Their movement can be purely random, or inhibited or enhanced by other molecules in the tissue. 
For example, a morphogen could bind to a receptor which would hinder its movement 
through the tissue. Morphogens can also be advected by tissue that is growing or moving. 
Cell based transport mechanisms include transcytosis \cite{dierick1998functional} 
and cytonemes \cite{Michael:RK99}. 

According to the transcytosis model, morphogens are taken up into the cell by 
endocytosis and are then released by exocytosis, 
facilitating their entry into a neighbouring cell \cite{rodman1990endocytosis}. 
In this way, morphogens can move through the tissue. The 
morphogen Decapentaplegic (Dpp) was proposed to spread by transcytosis in 
the \textit{Drosophila} wing disc, see \cite{Michael:ESG00}. 
However, transport by transcytosis would be too slow to explain the kinetics of 
Dpp spreading, cp.\ \cite{lander2002morphogen}, 
and further experiments refuted the transcytosis mechanism for Dpp 
transport in the wing disc, cf.\ \cite{Michael:SSSR+11}.

According to the cytoneme model, cytonemes, i.e.\ filapodia-like cellular projections, 
emanate from target cells to contact morphogen producing cells and 
\textit{vice versa}, cf.\ \cite{Michael:RK99}.
Morphogens are then transported along the cytonemes to the target cell. 
Several experimental studies support a role of cytonemes in morphogen 
transport across species, see \cite{Michael:KS14,Michael:BSAR+13,sanders2013specialized}.
 However, so far, the transport kinetics and the mechanistic 
details of the intracellular transport are largely unknown, and a validated theoretical 
framework to describe cytoneme-based transport is still missing. 
Accordingly, the standard transport mechanism in computational 
studies still remains diffusion. In this book chapter, we will only consider diffusion- 
and advection-based transport mechanisms.

\begin{figure}[htb]
\begin{center}
\includegraphics{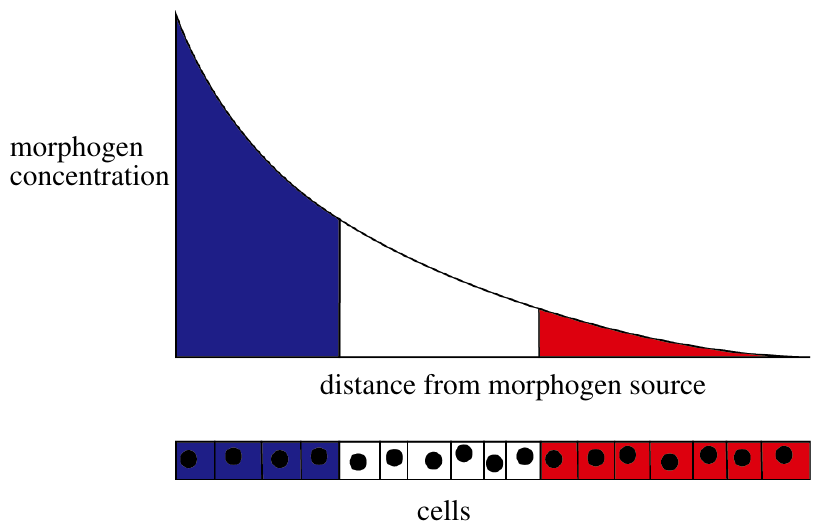}
	\end{center}
  \caption{Lewis Wolpert's French Flag Model}
  \label{fig:flag}
\end{figure}   
%\begin{figure}[htb]
%\begin{center}
%	\begin{tikzpicture}
%		\draw (1,0.25) node{\includegraphics[scale = 0.5]{./pics/French_Flag_blank_exp.png}};
%		\draw[align=left] (-3.1,1.1) node{morphogen\\[-0.4ex] concentration};
%		\draw (1.2,-1.2) node{distance from morphogen source};
%		\draw (0.5,-2.45) node {cells};
%	\end{tikzpicture}
%	\end{center}
%  \caption{Lewis Wolpert's French Flag Model}
%  \label{fig:flag}
%\end{figure}   

A key concept for morphogen-based patterning is Lewis Wolpert's 
\textit{French Flag model}, cf.\ \cite{wolpert1969positional}. 
According to the French Flag model, morphogens diffuse from a 
source and form a gradient across a tissue such that cells close to the 
source experience the highest morphogen concentration, 
while cells further away experience lower concentrations (Figure~\ref{fig:flag}). 
To explain the emergence of patterns such as the digits in the limb, 
Wolpert proposed that the fate of a tissue segment depends 
on whether the local concentration is above or below a patterning threshold. 
Thus, the cells with the highest concentration of the morphogen (blue) 
differentiate into one type of tissue, cells with a medium amount (white) 
into another and cells with the lowest concentration (red) into a third type, 
see Figure~\ref{fig:flag}. With the arrival of quantitative data, 
aspects of the French Flag model had to be modified, 
but the essence of the model has stood the test of time. 

In the original publication, the source was included as a fixed boundary condition. 
No reactions were included in the domain, 
but morphogen removal was included implicitly by including an absorbing 
(zero concentration) boundary condition on the other side. 
The resulting steady state gradient is linear and scales with the size of the domain, 
i.e.\ the relative pattern remains the same independent of the size of the domain. 
This is an important aspect as the patterning processes are typically robust to (small) 
differences in embryo size. 
Quantitative measurements have since shown that morphogen gradients are of exponential 
rather than linear shape, see \cite{gregor2005diffusion}. 
The emergence of exponentially shaped gradients can be explained with morphogen 
turn-over in the tissue, cp.\ \cite{lander2002morphogen}. 
However, such steady-state exponential gradients have a fixed length scale and thus 
do not scale with a changing length of the patterning domain. Scaled steady-state 
patterns would require the diffusion coefficient, the reaction parameters, or the flux, 
to change with the domain size, cf.\ \cite{umulis2013mechanisms,umulis2009analysis}. 
At least, an appropriate change in the diffusion coefficient can be ruled out \cite{gregor2005diffusion}. 
Intriguingly, pattern scaling is also observed on growing domains \cite{wartlick2011dynamics}. 
The observed dynamic scaling of the Dpp gradient can be explained with the pre-steady state kinetics 
of a diffusion-based transport mechanism (rather than the steady state gradient shape) 
and thus does not require any changes in the parameter values \cite{fried2014dynamic}. 
Finally, the quantitative measurements showed that the Dpp gradient amplitude increases continuously 
\cite{wartlick2011dynamics}.
A threshold-based read-out as postulated by the French Flag model is 
nonetheless possible because the amplitude increase and the imperfect 
scaling of the pre-steady state gradient compensate such that the Dpp concentration 
remains constant in the region of the domain where the Dpp-dependent pattern is defined, see \cite{fried2015read}. 
In summary, current experimental evidence supports a French Flag-like 
mechanism where tissue is patterned by the threshold-based read-out of morphogen gradients. 
However, these gradients are not necessarily in steady state. Accordingly, 
dynamic models of morphogen gradients must be considered on growing domains. 
To do so, a mathematical formalism and simulations are required. 

\subsection{Mathematical Description of Diffusing Morphogens}
Morphogen behaviour can be modelled mathematically using the reaction diffusion equation, 
which we derive here. We assume, for the moment, 
that there is no tissue growth and the movement of the morphogen is a consequence of random motion. 
We denote the concentration of a morphogen in the domain $\Omega\subset\mathbb{R}^d$ as $c({\bf{x}},t)$, 
as it is dependent on time and its spatial position in the domain. Then the total concentration in 
$\Omega$ is $\int_{\Omega} c({\bf{x}},t) \operatorname{d}\!{\bf{x}}$ and the rate of change of the total concentration is
\begin{equation}
\frac{\text{d}}{\text{dt}}\int_{\Omega} c({\bf{x}},t) \operatorname{d}\!{\bf{x}}
\label{eqn:conc_change}
\end{equation}
The rate of change of the total concentration in $\Omega$ is a result of interactions between the morphogens that impact their 
concentration and random movement of the morphogens. 
The driving force of diffusion is a decrease in Gibbs free energy or chemical potential difference. 
This means that a substance will generally move from an area of high concentration to an area of low concentration. 
The movement of $c({\bf{x}},t)$ is called the flux, i.e.\ the 
amount of substance that will flow through a unit area in a unit time interval. 
As the movement of the morphogen is assumed to be random, Fick's first law holds. 
The latter states that the magnitude of the morphogens movement from an area of high concentration
to one of low concentration is proportional to that of the difference between the concentrations, 
or concentration gradient, i.e.\
\begin{equation}
{\bf{j}} = -D \nabla c({\bf{x}},t),
\label{eqn:fick}
\end{equation}   
where \(D\) is the diffusion coefficient or diffusivity of the morphogen. 
This is a measure of how quickly the morphogen moves from a region of high 
concentration to a region of low concentration. 
The total flux out of $\Omega$ is then
\begin{equation*}
\int_{\partial\Omega} {\bf{j}} \cdot {\bf{n}} \ dS,
\end{equation*}
where $dS$ is the boundary of $\Omega$ and ${\bf{n}}$ is the normal vector to the boundary. 
Reactions between the morphogens also affect the rate of change of $c({\bf{x}},t)$. 
We denote the reaction rate $R(c)$. The rate of change of the concentration in the 
domain $\Omega$ due to morphogen interactions is
\begin{equation*}
\int_{\Omega} R(c) \operatorname{d}\!{\bf{x}}.
\end{equation*}
As the rate of change of the total concentration in $\Omega$ is the sum of the 
rate of change caused by morphogen interactions 
and the rate of change caused by random movement, we have
\begin{equation}
\frac{\text{d}}{\text{dt}}\int_{\Omega} c({\bf{x}},t) \operatorname{d}\!{\bf{x}} 
= - \int_{\partial\Omega} {\bf{j}} \cdot {\bf{n}} \ dS + \int_{\Omega} R(c) \operatorname{d}\!{\bf{x}}.
\label{eqn:balance}
\end{equation}
Now, the Divergence Theorem yields
\begin{equation}
\int_{\partial\Omega} {\bf{j}} \cdot {\bf{n}} \ dS = \int_{\Omega} \nabla \cdot {\bf{j}} \  
\operatorname{d}\!{\bf{x}}.
\label{eqn:div_thm}
\end{equation}
Substituting \eqref{eqn:div_thm} and \eqref{eqn:fick} into \eqref{eqn:balance} and 
exchanging the order of integration and differentiation using Leibniz's theorem gives
\begin{equation*}
\int_{\Omega} \frac{{\text{$\partial$c}}}{\text{$\partial$t}} - D\Delta c - R(c)  \operatorname{d}\!{\bf{x}} = 0.
\end{equation*}
Taking into account that this equilibrium holds for any control volume \(V\subset\Omega\), we obtain
the classical reaction-diffusion equation
\begin{equation}
 \frac{\text{$\partial$c}}{\text{$\partial$t}} = D\Delta c + R(c). 
 \label{eqn:RD}
\end{equation}

This partial differential equation (PDE) can be solved on a continuous domain to study 
the behaviour of morphogens in a fixed domain over time. If there is more than one morphogen 
then their respective concentrations
can be labelled $c_i({\bf x},t)$ for $i=1,\ldots,N$ where $N$ is the number of morphogens. 
The reaction term then describes the morphogen interactions i.e.\ \(R=R(c_1,\ldots,c_N)\). 
This results in a coupled system of PDEs, which, depending on the reaction terms, can be nonlinear. 
The accurate solution of these equations can be difficult and computationally costly. 

It is important to keep in mind that reaction-diffusion equations only describe the average behaviour of a diffusing substance. 
This approach is therefore not suitable if the number of molecules is small. In that case stochastic effects dominate, and stochastic, 
rather than deterministic, techniques should be applied, see \cite{gillespie2007stochastic, berg1993random, gillespie1977exact}.

\subsection{Morphogen Dynamics on Growing Domains}
\begin{figure}[htb]
\begin{center}
\includegraphics[scale=0.5]{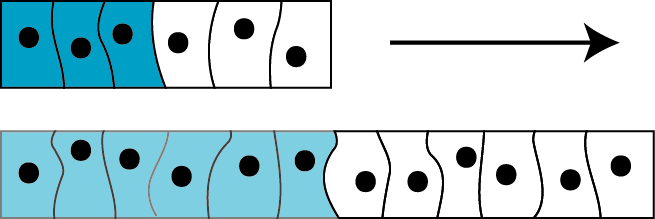}
\end{center}
\caption{Advection and dilution induced by a growing domain. 
The direction of growth is indicated by the arrow. 
Movement of the cells or tissue resulting from growth causes the morphogens to move (advection). 
Simultaneously cell division, or growth, dilutes morphogen molecules.}
\label{fig:adv_dil}
\end{figure}

In the previous paragraph we introduced morphogen dynamics on a fixed domain. 
However, tissue growth plays a key role in morphogenesis 
and can play a crucial part in the patterning process of the organism 
\cite{kondo1995reaction, henderson2002mechanical}. Growth can affect the distribution of the morphogens, 
transporting them via advection and impacting the concentration via dilution, see Figure~\ref{fig:adv_dil}. 
In turn, morphogens can influence tissue shape change and growth, 
for example, by initiating cell death and cell proliferation respectively. 
This results in a mutual feedback between tissue growth and morphogen concentration. 

%\begin{figure}[htb]
%\begin{center}
%\includegraphics[scale=0.5]{./pics/adv_dil.png}
%\end{center}
%\caption{Advection and dilution induced by a growing domain. 
%The direction of growth is indicated by the arrow. 
%Movement of the cells or tissue resulting from growth causes the morphogens to move (advection). 
%Simultaneously cell division, or growth, dilutes morphogen molecules.}
%\label{fig:adv_dil}
%\end{figure}

It is then necessary to modify equation \eqref{eqn:RD} to account for growth.
Applying Reynolds transport theorem to the left-hand side of equation \eqref{eqn:balance} we get
\begin{equation*}
\frac{\text{d}}{\text{dt}}\int_{\Omega_t} c({\bf{x}},t) \operatorname{d}\!{\bf{x}} 
= \int_{\Omega_t} \bigg( \frac{\partial c}{\partial t} 
+ \operatorname{div}(c{\bf{v}}) \bigg) \operatorname{d}\!{\bf{x}}.
\end{equation*}
For a more detailed derivation, we refer to \cite{Michael:ITFG+14}.
 This results in the reaction-diffusion equation on a growing domain:
\begin{equation}\label{eq:rdgrowth}
\frac{\text{$\partial$c}}{\text{$\partial$t}} + \operatorname{div}(c{\bf{v}}) = D\Delta c + R(c).
\end{equation}

By the Leibniz rule, there holds $\operatorname{div}(c{\bf{v}})=c\operatorname{div}({\bf{v}})+{\bf{v}}\cdot \nabla c$. 
These terms can be interpreted as the \textit{dilution}, i.e.\ the reduction in concentration of a solute in a solution, 
usually by adding more solvent, and \textit{advection}, 
i.e.\ movement of a substance in a fluid caused by the movement of the fluid, respectively.

\subsection{Modelling tissue growth}
The details of the process of tissue growth still remain to be elucidated. 
It is therefore an open question of how best to incorporate it into a model. 
One approach considers the velocity field to be dependent on morphogen concentration, 
i.e.\ ${\bf{v}}(c,{\bf{x}},t)$, where $c$ is again the morphogen concentration present in the tissue 
\cite{iber_control_2013, menshykau_kidney_2013}. 
Another is ``prescribed growth'', 
in which the velocity field ${\bf{v}}({\bf{x}},t)$ of the tissue is specified 
and the initial domain is moved according to this velocity field. 
A detailed measurement of the velocity field can be obtained from experimental data. 
To this end, the tissue of interest can be stained and imaged at sequential developmental time points. 
These images can then be segmented to determine the shape of the domain. 
Displacement fields can be calculated by computing the distance between the domain boundary 
of one stage and that of the next. A velocity field in the domain can then be interpolated, 
for example by assuming uniform growth between the centre of mass and the nearest boundary point. 
To this extent, high quality image data is required 
to enable detailed measurements of the boundary to be extracted. 
For a detailed review on this process see \cite{gomez2017image, iber_image-based_2015}.

There are also other techniques to model tissue growth. If the local growth rate of the tissue is known,
the Navier-Stokes equation can be used. 
Tissue is assumed to be an incompressible fluid and tissue growth can then 
be described with the Navier-Stokes equation for incompressible flow of Newtonian fluids, which reads
\begin{equation}\label{eq:NS}
\begin{aligned}
\rho \big( \partial_t {\bf{v}} +({\bf v}\cdot\nabla){\bf{v}}\big) &= -\nabla p + \mu \bigg(\Delta {\bf{v}} + \frac{1}{3} \nabla (\operatorname{div}{\bf v}) \bigg) +{\bf f},\\
\rho \operatorname{div} {\bf{v}} &= \omega S,
\end{aligned}
\end{equation}
where $\rho$ is fluid density, $\mu$ dynamic viscosity, $p$ internal pressure, 
${\bf f}$ external force density and ${\bf v}$ the fluid velocity field. 
The term $\omega S$ is the local mass production rate. The parameter $\omega$ 
is the molecular mass of the cells. 
The impact of cell signalling on growth can be modelled by having the source 
term $S$ dependent on the morphogen concentration,
i.e.\ \(S=S(c)\). Note that the source term results in isotropic growth. 
External forces as implemented in the ${\bf f}$ term can induce anisotropic growth. 
Based on measurements, the Reynolds number for tissues is typically small, e.g.\ in embryonic
tissue it is assumed to be of the order $10^{-14}$, see \cite{Michael:ITFG+14}. 
Accordingly, the terms on the left-hand side of Equation~\eqref{eq:NS} can be neglected, as in Stokes' Flow. 

The Navier-Stokes description, with a source term dependent on signalling, 
has been used in simulations of early vertebrate limb development, see \cite{dillon2003short}. 
An extended anistropic formulation has been applied to \textit{Drosophila} 
imaginal disc development in \cite{bittig2008dynamics}. 
It has also been used to model bone development, cf.\ \cite{tanaka2013inter}, 
and coupled with a travelling wave to simulate the developing 
\textit{Drosophila} eye disc, see \cite{Michael:FSAL+16}. 
In the case of the developing limb, the proliferation rates were later determined, 
see \cite{boehm2010role}. 
They were then used as source terms in the isotropic Navier-Stokes tissue model. 
There was, however, a significant discrepancy between the predicted and actual growth. 
The shapes of the experimental and simulated developing limb were qualitatively 
different and the actual expansion of the limb was much larger than expected. 
This shows that limb expansion must result from anisotropic processes, and suggested 
that the growth of the limb could in part be due to cell migration rather 
than solely local proliferation of cells.

\subsection{Tissue mechanics}\label{sec:TissueMech}
%%%%%%%%%%%%%%%%%%%%%%%%%%%%%%%%%%%%%%%%%%%%%%%%%%%%%%%%%%%%%%%%%%%%%%%%%%%%%%%%%%%%%%%%%
%\input{ContinuumApproaches}
\tikzset{pics/.cd,
grid/.style args={(#1)#2(#3)#4(#5)#6(#7)#8}{code={%
\tikzset{pics/grid/dimensions=#8}%
% still don't get what the template does, so I fixed coordinates
% the corners of the grid are now a b c d
\coordinate (a) at (#1);
\coordinate (b) at (#3);
\coordinate (c) at (#5);
\coordinate (d) at (#7);
\foreach \i in {0,...,\y}
  \draw [pic actions/.try] ($(a)!\i/\y!(d)$) -- ($(b)!\i/\y!(c)$);
\foreach \i in {0,...,\x}
  \draw [pic actions/.try] ($(a)!\i/\x!(b)$) -- ($(d)!\i/\x!(c)$);

\path (#1) coordinate (-1) (#3) coordinate (-2)
      (#5) coordinate (-3) (#7) coordinate (-4);
}},
grid/dimensions/.code args={#1x#2}{\def\x{#1}\def\y{#2}}
}
\tikzset{pics/.cd,
prtcl/.style args={(#1)#2(#3)#4(#5)#6(#7)#8}{code={%
\tikzset{pics/prtcl/dimensions=#8}%
% still don't get what the template does, so I fixed coordinates
% the corners of the grid are now a b c d
\coordinate (a) at (#1);
\coordinate (b) at (#3);
\coordinate (c) at (#5);
\coordinate (d) at (#7);

\foreach \i in {1,...,\x} {
	\foreach \j in {1,...,\y} {
	\pgfmathsetmacro\weightx{(\i-0.5)/\x}
	\pgfmathsetmacro\weightxx{1-\weightx}
	\pgfmathsetmacro\weighty{(\j-0.5)/\y}
	\pgfmathsetmacro\weightyy{1-\weighty}
		\fill[pic actions/.try] ($\weighty * \weightx * (a)
						+\weighty * \weightxx*(b) 
						+ \weightyy * \weightx * (d)
						+ \weightyy * \weightxx * (c)$) circle [radius=2pt];
		}
		}
}},
prtcl/dimensions/.code args={#1x#2}{\def\x{#1}\def\y{#2}}
}
%\subsection{Continuum approaches}
Tissue expands and deforms during growth. Given its elastic properties, 
stresses must emerge in an expanding and deforming tissue. 
Cell rearrangements are able to dissipate these stresses and 
numerous experiments confirm the viscoelastic properties of tissues 
\cite{Michael:For98,Michael:FFSS98,Michael:FFPS94,Michael:MMKA+01}. 
Over long time scales, as characteristic for many developmental processes, 
tissue is therefore typically represented as a liquid, viscous material and is 
then described by the Stokes equation \cite{Michael:DO99,Michael:FSAL+16,Michael:ITFG+14}. 
Over short time scales, however, tissues have mainly elastic properties. 
Continuum mechanical models are widely used to simulate the mechanical properties of tissues, 
see e.g.\ \cite{Michael:Fun93,Michael:Tab04} and the references therein.
Continuum mechanical descriptions usually consist of three parts:
the \textit{kinematics}, which describes the motions of objects, 
their displacements, velocity and acceleration, the \textit{constitutive equations}, 
which model the material laws and describe the response of the material 
to induced strains in terms of stress and the underlying \textit{balance principle}, 
which describes the governing physical equations. 
For a comprehensive introduction into continuum mechanics, 
we refer to \cite{Michael:Cia88,Michael:Hol00}.

\begin{figure}
\begin{center}
\includegraphics[scale=0.45]{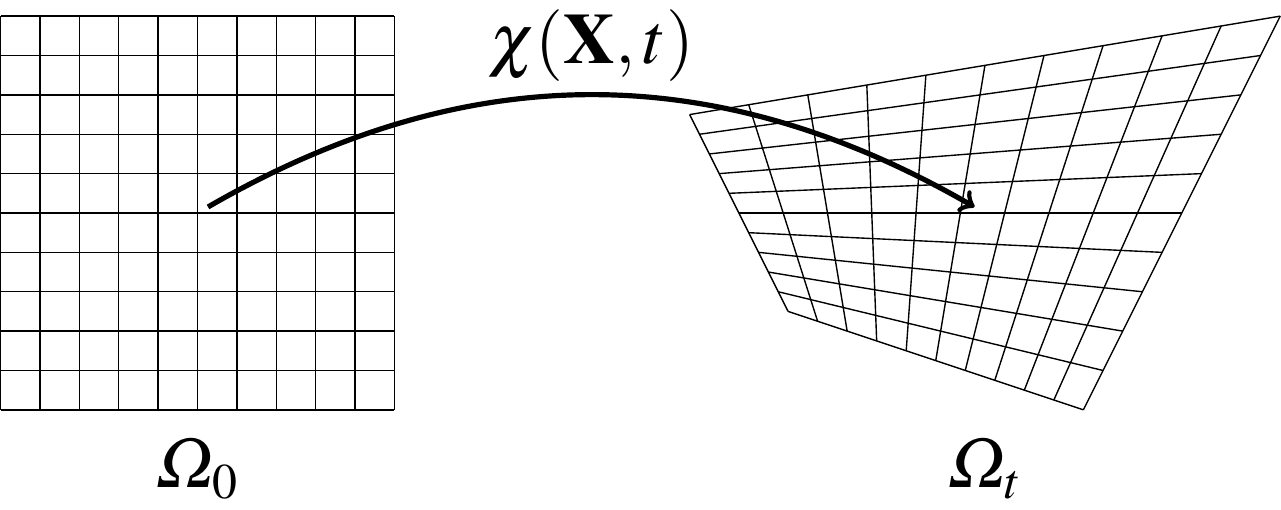}
\end{center}
\caption{\label{fig:defField}Deformation of the domain \(\Omega_0\) via the deformation field \(\boldsymbol{\chi}\).}
\end{figure}
%\begin{figure}
%\begin{center}
%\scalebox{0.45}{\begin{tikzpicture}
%\pic at (0,0) [black] {grid={(0,0) (4,0)  (4,4) (0,4)  10x10}};
%\pic at (8,1) [black]  {grid={(0,0) (3,-1) (5,3) (-1,2) 10x10}};
%\draw (2,2)node(N4){};
%\draw (10,2)node(N5){};
%\path[->,line width=1.5pt]  (N4) edge [bend left] (N5);
%\draw (6,3.7)node{\huge$\boldsymbol{\chi}({\bf X},t)$};
%\draw (2,-0.6)node{\huge$\Omega_0$};
%\draw (10,-0.6)node{\huge$\Omega_t$};
%\end{tikzpicture}}
%\end{center}
%\caption{\label{fig:defField}Deformation of the domain \(\Omega_0\) via the deformation field \(\boldsymbol{\chi}\).}
%\end{figure}

The mathematical representation of kinematics, i.e.\ the description of motion of points and bodies, 
is usually performed with respect to two different frameworks. They are called \textit{Lagrangian} 
(or \textit{material}) and \textit{Eulerian} (or \textit{spatial}) coordinates.
The Lagrangian framework adopts a particle point of view, for example the perspective of a single cell, 
and tracks its movement over time.
In contrast to this, the Eulerian framework adopts the perspective of an entire body, 
for example a tissue, and describes its position over time
with respect to a given coordinate frame. 
More precisely, let \(\Omega_0\subset\mathbb{R}^d\) denote a body and 
\[
\boldsymbol{\chi}\colon\Omega_0\times[0,\infty)\to\mathbb{R}^d,
\quad ({\bf X},t)\mapsto\boldsymbol{\chi}({\bf X},t)
\] 
a \textit{deformation field}. The deformed body at a given time \(t\in[0,\infty)\) is then denoted by 
\(\Omega_t\isdef\boldsymbol{\chi}(\Omega_0,t)\), see Figure\ \ref{fig:defField} for a visualisation.
The position of the particle \({\bf X}\in\Omega_0\) at time \(t\geq0\) is therefore given by 
\({\bf x}=\boldsymbol{\chi}({\bf X},t)\), which is the description in Eulerian coordinates.
On the other hand, we can also consider \({\bf X}=\boldsymbol{\chi}^{-1}({\bf x},t)\), 
which is the description in Lagrangian coordinates.
More important than the deformation is the \textit{displacement}
\[
{\bf U}({\bf X},t)\isdef {\bf x}({\bf X},t)-{\bf X}\quad\text{or}\quad{\bf u}({\bf x},t)\isdef{\bf x}-{\bf X}({\bf x},t),
\]
respectively. Based on the displacement, one can consider balance principles of the form 
\[
\operatorname{div}\boldsymbol{\sigma}+{\bf f} = \rho\ddot{\bf u},
\]
which is Newton's second law and is also known as Cauchy's first equation of motion, see e.g.\ \cite{Michael:Hol00}.
Herein, the tensor field \(\boldsymbol\sigma\) characterises the stresses inside the body, 
the vector field \({\bf f}\) summarises internal forces, \(\rho\) denotes
the mass density and \(\ddot{\bf u}\) is the acceleration. Thus, at steady state, the equation simplifies to
\[
-\operatorname{div}\boldsymbol{\sigma}={\bf f}.
\]
Note that the steady state in morphogen concentrations is reached very fast compared to the time scale 
on which growth happens.

Several models exist to describe material behaviour. 
A material is called \textit{elastic} if there exists a \textit{response function} 
with \[\boldsymbol{\sigma}={\bf g}({\bf F}),\] where 
 \[
 {\bf F}({\bf X},t)\isdef\nabla\boldsymbol{\chi}=[\partial_{X_j}\chi_i]_{i,j}
 \] 
 is the \textit{deformation gradient}. 
 \textit{Linearly elastic} materials are described by \textit{Hooke's law}. 
 In this case, the function \({\bf g}\) is linear.
 For the description of tissues, non-linear material responses are better suited. 
 To that end, \textit{hyperelastic} material models are used. They are characterised by the
 response function
\[
{\bf g}({\bf F})=J^{-1}\frac{\partial W({\bf F})}{\partial{\bf F}}{\bf F}^\intercal,
 \]
 where \(J\isdef\det{\bf F}\) and \(W\) is a scalar \textit{strain energy density function}. 
 For the modelling of soft tissues, \textit{Fung-elastic} materials might be employed, 
 see e.g.\ \cite{Michael:Fun93,Michael:Tab04}. Here the strain energy density function is
 for example given by
 \[
 W=\frac{C}{\alpha}\bigg[e^{\alpha(I_1-3)}-1\bigg],\quad\text{with }C,\alpha>0,
 \quad I_1\isdef\operatorname{trace}({{\bf FF}^\intercal}).
 \]
 
  \begin{figure}[htb]
\begin{center}
\includegraphics{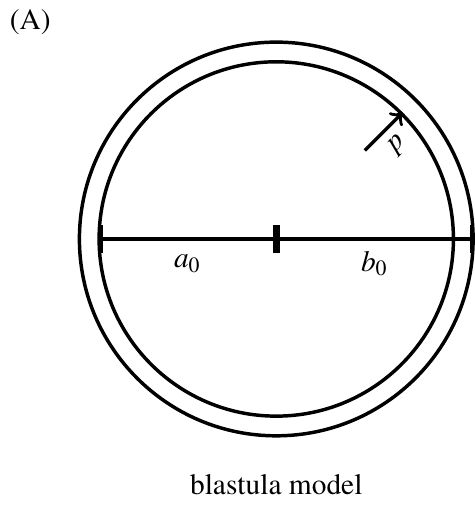}\hfill
\includegraphics{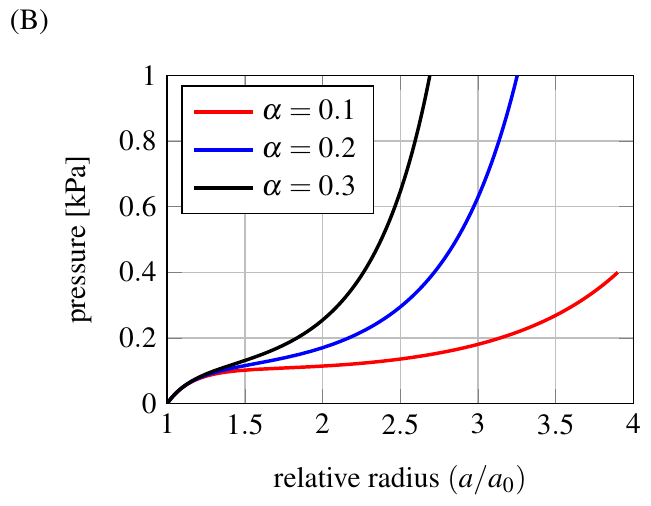}
\end{center}
\caption{\label{fig:Blastula}Numerical simulation of the blastula. (A) Model of the sea urchin blastula. 
					 (B) Pressure versus radius curves for different values of the parameter \(\alpha\).}
\end{figure}

In \cite{Michael:Tab04}, this model is suggested to simulate the blastula stage of the sea urchin.
Figure \ref{fig:Blastula}.A shows the corresponding computational model. 
By considering only a cross section, the model can be reduced to two spatial dimensions. 
In Figure~\ref{fig:Blastula}.B, pressure versus radius curves 
for different values of the parameter \(\alpha\) are shown, where we assume 
that an interior pressure \(p\) acts on the interior wall. 
As can be seen, the material stiffens for increasing values of \(\alpha\).
For the numerical simulations, the thickness of the blastula is set to 
\(b_0-a_0=75-50=25\) and we chose \(C=0.2\text{[kPa]}\), see 
 also \cite{Michael:Tab04,Michael:IP17}.
 %%%%%%%%%%%%%%%%%%%%%%%%%%%%%%%%%%%%%%%%%%%%%%%%%%%%%%%%%%%%%%%%%%%%%%%%%%%%%%%%%%%%%%%%%
%\input{CellbasedSimulation}
\section{Cell-based simulation frameworks}\label{sec:CellBased}

\begin{figure}[htb]
\begin{center}
\includegraphics{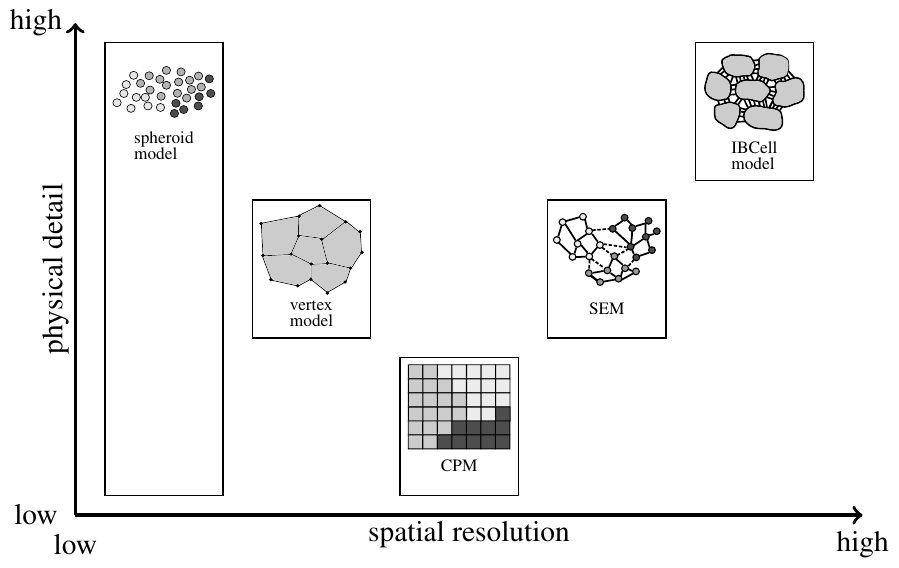}
\end{center}
\caption{\label{fig:cellbased}Cell-based simulation frameworks and 
their arrangement with respect to physical detail and spatial resolution.}
\end{figure}
%\begin{figure}[htb]
%\begin{center}
%\begin{tikzpicture}
%\draw[->,line width=1pt] (0,0) -- (0,5);
%\draw[->,line width=1pt] (0,0) -- (8,0);
%\draw (0.3,0.2) rectangle (1.5,4.8);
%\draw (1.8,1.8) rectangle (3,3.2);
%\draw (3.3,0.2) rectangle (4.5,1.6);
%\draw (4.8,1.8) rectangle (6,3.2);
%\draw (6.3,3.4) rectangle (7.5,4.8);
%\draw (0.9,4.3) node{\includegraphics[scale=.09, clip, trim= 0 0 0 0]{./pics/spheroid.pdf}};
%\draw (2.4,2.7) node{\includegraphics[scale=.25, clip, trim= 0 0 0 0]{./pics/vertex.pdf}};
%\draw (3.9,1.1) node{\includegraphics[scale=.312, clip, trim= 0 0 0 0]{./pics/cpm.pdf}};
%\draw (5.4,2.7) node{\includegraphics[scale=.62, clip, trim= 0 0 0 0]{./pics/sem.pdf}};
%\draw (6.9,4.3) node{\includegraphics[scale=.6, clip, trim= 0 0 0 0]{./pics/ibm.pdf}};
%
%%\draw[fill=white, white] (6.8,4.6) rectangle (7.3,4.75);
%\draw[align=left] (0.9,3.75) node{\tiny spheroid\\[-1.4ex] \tiny model};
%\draw[align=left] (2.4,2.05) node{\tiny vertex\\[-1.4ex] \tiny model};
%\draw[align=left] (3.9,0.5) node{\tiny CPM};
%\draw[align=left] (5.4,2.1) node{\tiny SEM};
%\draw[align=left] (6.9,3.65) node{\tiny IBCell\\[-1.4ex]\tiny model};
%\draw (0,-0.3) node{low};
%\draw (8,-0.3) node{high};
%\draw (-0.4,0) node{low};
%\draw (-0.4,5) node{high};
%\draw (4,-0.2) node{spatial resolution};
%\draw (-0.2,2.5) node[rotate=90]{physical detail};
%\end{tikzpicture}
%\end{center}
%\caption{\label{fig:cellbased}Cell-based simulation frameworks and 
%their arrangement with respect to physical detail and spatial resolution.}
%\end{figure}

Cell-based simulations complement continuum models, 
and are important when cellular processes need to be included explicitly, 
i.e.\ cell adhesion, cell migration, cell polarity, cell division, and cell differentiation. 
Cell-based simulations are particularly valuable when local rules 
at the single-cell level give rise to global patterns. 
Such emergence phenomena have been studied with agent-based models in a range of fields, 
long before their introduction to biology. \
Early agent-based models of tissue dynamics were simulated on a lattice, 
and each cell was represented by a single, autonomous agent that moved and interacted 
with other agents according to a set of local rules. The effects of secreted morphogens 
or cytokines can be included by coupling the agent-based model with reaction-diffusion 
based continuum models, as done in a model of the germinal center reaction during an immune 
response \cite{Meyer-Hermann2006}. 
In this way, both direct cell-cell communication and long-range interactions can be realised. 

A wide range of cell-based models has meanwhile been developed. 
The approaches differ greatly in their resolution of the underlying physical 
processes and of the cell geometries, and have been realised both as lattice-based 
and lattice-free models. In lattice-based models the spatial domain is 
represented by a one, two or three dimensional lattice and a cell occupies a certain number of lattice sites. 
Cell growth can be included by increasing the number of lattice sites per cell and cell proliferation 
by adding new cells to the lattice. In off-lattice approaches, cells can occupy an unconstrained area in the domain.
Similarly to on-lattice models, tissue growth can be implemented by modelling cell growth and proliferation.
However, cell growth and proliferation is not restricted to discrete lattice sites. 
Among the off-lattice models it can be distinguished between center-based models, 
which represent cell dynamics via forces acting on the cell centers, and deformable 
cell models that resolve cell shapes. In case of a higher resolution of the cell geometry, 
the cell-based models can be coupled to signalling models, where components are restricted 
to the cells or the cell surface. A higher resolution and the independence from a lattice 
permit a more realistic description of the biological processes, 
but the resulting higher computational costs limit the tissue size and time frame that can be simulated. 

In the following, we will provide a brief overview of the most widely used 
cell-based models for morphogenetic simulations, 
i.e.\ the \textit{Cellular Potts model}, the \textit{spheroid model},
the \textit{Subcellular Element model}, the \textit{vertex model}, 
and the \textit{Immersed Boundary Cell model}, see Figure~\ref{fig:cellbased} 
for their arrangement with respect to physical detail and spatial resolution.
In the following, we will focus on the main ideas behind each model and name common 
software frameworks that implement the aforementioned methods. 
A more detailed description of the models and their applications in 
biology can be found in \cite{Merks2009,Tanaka2015,VanLiedekerke2015}.

The \textit{Cellular Potts Model} (CPM) is a typical on-lattice approach 
as it originates from the Ising model, see \cite{Ising1925}. 
It represents the tissue as a lattice where each lattice site carries 
a spin value representing the cell identity. 
The update algorithm of the CPM is the Metropolis algorithm, 
cf.\ \cite{Metropolis1953},
which aims to minimize the Hamiltonian energy function which is defined over the entire lattice. 
In the original CPM introduced in \cite{Graner1992},
the Hamiltonian includes a volume constriction term and a cell-cell adhesion term:
\begin{equation*}
  H = \sum_{\sigma} \lambda_{v}(V_{\sigma} - V_{\sigma}^{T})^{2} +
  \sum_{(\bf{x},\bf{x}')} J\Big( \tau\big(\sigma(\bf{x})\big) ,  
  \tau\big(\sigma(\bf{x}')\big)\Big) 
  \cdot
  \Big(1-\delta\big(\sigma(\bf{x}),\sigma( \bf{x}')\big) \Big).
\end{equation*}
The first term describes the volume constriction with 
$\lambda_{v}$ being the coefficient controlling the energy penalisation, 
$V_{\sigma}$ being the actual volume and $V_{\sigma}^{T}$ the target volume of cell $\sigma$.
The second term represents the cell-cell adhesion, 
where $J\left(\tau\left(\sigma\left(\bf{x}\right)\right),  
\tau\left(\sigma\left(\bf{x}'\right)\right)  \right)$ 
denotes the surface energy term between two cell types 
and $\delta(\cdot,\cdot)$ the Dirac \(\delta\)-function. 
The CPM has been used to simulate various processes in morphogenesis, 
including kidney branching morphogenesis \cite{Hirashima:2009er}, 
somitogenesis \cite{Hester:2011cc}, and chicken limb development \cite{Manuscript2009}. 
As for many lattice-based algorithms, 
an advantage of the CPM is the efficient application of high performance 
computing and parallelisation techniques, see e.g.\ \cite{Tanaka2015}. 
The main limitations of the CPM concern its high level of abstraction 
that limits the extent to which the simulations can be validated with experimental data:
the interpretation of the temperature in the Metropolis algorithm is not straightforward 
and there is no direct translation between iteration steps and time. 
Moreover, the representation of cells and cell growth is coarse, 
and biophysical properties are difficult to directly relate to measurements. 
The open-source software framework CompuCell3D, see \cite{Swat2012}, is based on the CPM.

The \textit{spheroid model} is an example for off-lattice \textit{agent-based models} 
and represents cells as particle-like objects being a typical off-lattice approach.
The cells are assumed to have a spherical shape being represented by a soft sphere 
interaction potential like the Johnson-Kendall-Roberts potential or the Hertz potential.
The evolution of cells in time in the spheroid model can be performed in two different ways: 
deterministically by solving the equation of motion 
\begin{equation*}
\eta \frac{\operatorname{d}\!\textbf{x}_i}{\operatorname{d}\!t}=\sum_{i} \bf{F}_i,
\end{equation*}
where $\eta$ is the mobility coefficient and $\bf{F}_i$ represent the forces acting on each particle, 
or by solving the stochastic Langevin equation. 
The simple representation of cells enables the simulation of a large number of cells with the 
spheroid model and further, to simulate tissues in 3D.
However, as all cells are represented equally, 
there are no cellular details represented and 
thus also the coupling of tissue dynamics to morphogen dynamics is restricted.
The 3D software framework CellSys, cf.\ \cite{Hoehme2010}, 
is built based on the spheroid model.

The \textit{Subcellular Element Model} (SEM) represents a cell by many subcellular elements 
assuming that the inner of a cell, i.e.\ the cytoskeleton, can be subdivided.
The elements are represented by point particles which interact via forces that are 
derived from interaction potentials such as the Morse potential.
A typical equation of motion for the position $\bf{y}_{\alpha_i}$ of a subcellular element 
$\alpha_i$ of cell $i$ reads \cite{Tanaka2015}
\begin{equation*}
  \eta \frac{\partial {\bf y}_{\alpha_{i}}}{\partial t} = \zeta_{\alpha_{i}}
  			- \nabla_{\alpha_{i}} \sum_{\beta_{i} \neq \alpha_{i}} V_{\text{intra}}(| {\bf y}_{\alpha_{i}} - {\bf y}_{\beta_{i}}|)
  			- \nabla_{\alpha_{i}} \sum_{j\neq i} \sum_{\beta_{j}} V_{\text{inter}}(| {\bf y}_{\alpha_{i}} - {\bf y}_{\beta_{j}}|)
\end{equation*}
with $\eta$ being the viscous damping coefficient and $\zeta_{\alpha_{i}}$ Gaussian noise.
The first term describes intra-cellular interactions between the subcellular element $\alpha_i$ and 
all other subcellular elements $\beta_i$ of cell $i$.
The second term represents inter-cellular interaction that takes into account all pair-interactions between 
subcellular elements $\beta_{j}$ of neighboring cells $j$ of cell $i$.
For intra-and inter-cellular interaction potentials $V_{\text{intra}}, V_{\text{inter}}$ 
for example the Morse potential can be used:
\begin{equation*}
V(r) = U_0 \exp \left(\frac{-r}{\epsilon_1}\right) - V_0 \exp \left(\frac{-r}{\epsilon_2}\right),
\end{equation*}
where $r$ is the distance between two subcellular elements, $U_0,V_0$ are the energy scale parameters and 
$\epsilon_1, \epsilon_2$ are the length scale parameters defining the shape of the potential.
Therefore, the SEM is very similar to agent-based models with the difference 
that each point particle represents parts of and not an entire cell.
The SEM offers an explicit, detailed resolution of the cell shapes, further, a 3D implementation is straightforward.
A disadvantage of the SEM is its high computational cost.

In the \textit{vertex model}, cells are represented by polygons, 
where neighboring cells share edges and an intersection point of edges is a vertex. 
It was first used in 1980 to study epithelial sheet deformations \cite{Honda1980}.
The movement of the vertices is determined by forces acting on them, 
which can either be defined explicitly, see e.g.\ \cite{Weliky1990}, 
or are derived from energy potentials, cf.\ \cite{Nagai2001}. 
A typical energy function has the following form, see \cite{Tanaka2015},
  \begin{equation*}
    E(\textbf{R}_i)=
    \sum_{\alpha}\frac{K_{\alpha}}{2}(A_{\alpha}-A_0)^2 +
    \sum_{\langle i,j\rangle}\Lambda_{ij}l_{ij}+
    \sum_{\alpha}\frac{\Gamma_{\alpha}}{2}L_{\alpha}^2
  \end{equation*}
with $\bf{R}_i$ being the junctions direction of vertex $i$.
The first term describes elastic deformations of a cell with $K_{\alpha}$ being the area elasticity coefficient,  
$A_{\alpha}$ the current area of a cell and $A_0$ the resting area.
The second term represents cell movements due to cell-cell adhesion via the line tension between 
neighboring vertices $i$ and $j$ with $\Lambda_{ij}$ being the line tension coefficient and $l_{ij}$ the edge length.
The third term describes volumetric changes of a cell via the perimeter contractility, 
with $\Gamma_{\alpha}$ being the contractility coefficient, $L_{\alpha}$ the perimeter of a cell, 
see \cite{Fletcher2013,Fletcher2014} and the references therein for further details.
Different approaches have been developed to move the vertices over time.
The explicit cell shapes in the vertex model allow for a relatively high level of detail, 
however, cell-cell junction dynamics, cell rearrangements etc.\ require a high level of abstraction. 
The vertex model is well suited to represent densely packed epithelial tissues, 
but there is no representation of the extracellular matrix.
Computationally, the vertex model is still relatively efficient.
The software framework Chaste, cp.\ \cite{Pitt-Francis2009,Mirams:2013it}, 
is a collection of cell-based tissue model implementations that includes the vertex model amongst others.

In the \textit{Immersed Boundary Cell Model} (IBCell model) the cell boundaries are discretized resulting 
in a representation of cells as finely resolved polygons.
These polygons are immersed in a fluid and, in contrast to the vertex model, 
each cell has its own edge, cf.\ \cite{Rejniak2007}.
Therefore, there are two different fluids: fluid inside the cells representing the cytoplasm 
and fluid between the cells representing the inter-cellular space.
The fluid-structure interaction is achieved as follows: 
iteratively, the fluid equations for intra- and extracellular fluids are solved, 
the velocity field to the cell geometries is interpolated, the cells are moved accordingly, 
the forces acting on the cell geometries are recomputed and distributed to the surrounding fluid, 
and the process restarts.
In other words, the moving fluids exert forces on the cell membranes, 
and the cell membranes in turn exert forces on the fluids.
Further, different force generating processes can be modeled on the cell boundaries, 
such as cell-cell junctions or membrane tensions for example by inserting 
Hookean spring forces between pairs of polygon vertices.
The IBCell model offers a high level of detail in representing the cells, down to individual cell-cell junctions.
Furthermore, the representation of tissue mechanics such as cell division or cell growth can be easily implemented.
A disadvantage of the IBCell model is its inherent computational cost.
The open source software framework LBIBCell, cp.\ \cite{Tanaka2015b}, is built on the combination 
of the IBCell model and the Lattice Boltzmann (LB) method, see Paragraph~\ref{ssec:LB}. 
LBIBCell realises the fluid-structure interaction is by an iterative algorithm. Moreover,
it allows for the coupled simulation of cell dynamics and biomolecular signaling.

\section{Overview of numerical approaches}\label{sec:NumApp}
%%%%%%%%%%%%%%%%%%%%%%%%%%%%%%%%%%%%%%%%%%%%%%%%%%%%%%%%%%%%%%%%%%%%%%%%%%%%%%%%%%%%%%%%%
As we have seen so far, the mathematical description of biological processes 
leads to complex systems of reaction diffusion equations, which might even be
defined with respect to growing domains. 
In this section, we give an overview of methods to solve these equations numerically. 
For the discretisation of partial differential equations, 
we consider the finite element method, 
which is more flexible when it comes to complex geometries than,
the also well known, finite volume and finite difference methods, 
see e.g.\ \cite{EGH00,Michael:Bra17} and the references therein. 
The numerical treatment of growing domains can be incorporated by either the 
arbitrary Lagrangian-Eulerian method or the Diffuse-Domain method. 
Finally, we consider the Lattice Boltzmann method, 
which is feasible for the simulation of fluid dynamics 
and the simulation of reaction diffusion equations.

\subsection{Finite element method}
The finite element method is a versatile tool to treat partial
differential equations in one to three spatial dimensions numerically.
The method is heavily used in practice to solve engineering, physical and biological problems. 
Finite elements were invented
in the 1940s, see the pioneering work \cite{Michael:Cou43}, and are textbook knowledge in the meantime, 
see e.g.\ \cite{Michael:Bra17,Michael:BS02,Michael:SB91,Michael:ZTZ13}. 
Particularly, there exists a wide range of commercial and open source software 
frameworks that implement the finite element method, e.g.\ 
\textit{COMSOL}\footnote{https://www.comsol.com}, 
\textit{dune-fem}\footnote{https://www.dune-project.org/modules/dune-fem} 
and \textit{FEniCS}\footnote{https://fenicsproject.org}.

The pivotal idea, the Ritz-Galerkin method, dates back to the
beginning of the 20th century, see \cite{Michael:GW12} for a historical overview.
The underlying principle is the fundamental lemma of calculus of variations: 
Let \(g\colon(0,1)\to\mathbb{R}\) be a continuous function.
If 
\[
\int_0^1 gv\operatorname{d}\!x = 0\quad\text{for all }v\in C^\infty_0(0,1),
\]
i.e.\ for all compactly supported and smooth functions \(v\) on (0,1), 
then there holds \(g\equiv 0\), cf.\ \cite{Michael:GF63}.
We can apply this principle to solve the second order boundary value problem
\begin{equation}\label{eq:1Dode}
-\frac{\operatorname{d}^2\!u}{\operatorname{d}\!x^2} = f\quad\text{in }(0,1)\quad\text{and}\quad u(0)=u(1)=0
\end{equation}
for a continuous function \(f\colon(0,1)\to\mathbb{R}\), numerically. 
The fundamental lemma of calculus of variations yields
\[
-\frac{\operatorname{d}^2\!u}{\operatorname{d}\!x^2} = f\quad\Leftrightarrow\quad\int_0^1\bigg(-\frac{\operatorname{d}^2\!u}{\operatorname{d}\!x^2}-f\bigg)v\operatorname{d}\!x=0\quad\text{for all }v\in C^\infty_0(0,1).
\]
Rearranging the second equation and integrating by parts then leads to
\[
-\frac{\operatorname{d}^2\!u}{\operatorname{d}\!x^2} = f\quad\Leftrightarrow\quad
\int_0^1\frac{\operatorname{d}\!u}{\operatorname{d}\!x}\frac{\operatorname{d}\!v}{\operatorname{d}\!x}\operatorname{d}\!x=
\int_0^1 fv\operatorname{d}\!x\quad\text{for all }v\in C^\infty_0(0,1).
\]
Dependent on the right hand side, the above equation does not necessarily have a solution \(u\in C^2(0,1)\). 
However, solvability is guaranteed in the 
more general function space \(H_0^1(0,1)\), 
which consists of all weakly differentiable functions with square integrable derivatives.
Then, introducing the bilinear form
\[
a\colon H^1_0(0,1)\times H^1_0(0,1)\to\mathbb{R},\quad a(u,v)\isdef\int_0^1\frac{\operatorname{d}\!u}{\operatorname{d}\!x}\frac{\operatorname{d}\!v}{\operatorname{d}\!x}\operatorname{d}\!x,
\]
and the linear form
\[
\ell\colon H^1_0(0,1)\to\mathbb{R},\quad \ell(v)\isdef\int_0^1 fv\operatorname{d}\!x,
\]
the \textit{variational formulation} of \eqref{eq:1Dode} reads
\[
a(u,v)=\ell(v)\quad\text{for all }v\in H^1_0(0,1).
\]

The idea of the \textit{Ritz-Galerkin method} is now, 
to look for the solution to the boundary value problem only in a finite dimensional subspace
\(V_N\subset H^1_0(0,1)\):
\begin{align*}
&\text{Find }u_N\in V_N\text{ such that}\\
&\quad a(u_N,v_N)=\ell(v_N)\quad\text{for all }v_N\in V_N.
\end{align*}
Let \(\varphi_1,\ldots,\varphi_N\) be a basis of \(V_N\). 
Then, there holds 
\(u(x)\approx\sum_{j=1}^Nu_j\varphi_j(x)\). 
Moreover, due to linearity, it is sufficient to consider only the basis 
functions \({\varphi}_i\) as \textit{test functions} \(v_N\in V_N\).
Thus, we obtain
\[
a\bigg(\sum_{j=1}^Nu_j\varphi_j,\varphi_i\bigg)=\sum_{j=1}^Na(\varphi_j,\varphi_i)u_j
=\ell(\varphi_i)\quad\text{for }i=1,\ldots,N
\]
and consequently, by setting \({\bf A}\isdef[a(\varphi_j,\varphi_i)]_{i,j=1}^N\in\mathbb{R}^{N\times N}\), 
\({\bf u}\isdef[u_i]_{i=1}^N\in\mathbb{R}^N\) and
\({\bf f}\isdef[\ell(\varphi_i)]_{i=1}^N\in\mathbb{R}^N\), we end up with the linear system of equations
\[
{\bf Au}={\bf f}.
\]
The latter can now be solved by standard techniques from linear algebra.

\begin{figure}[htb]
\begin{center}
\includegraphics{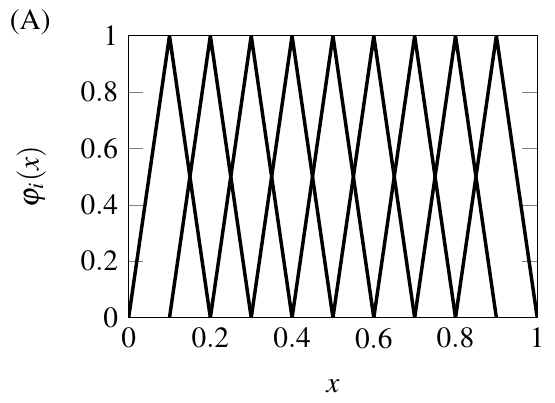}\hspace{1em}
\includegraphics{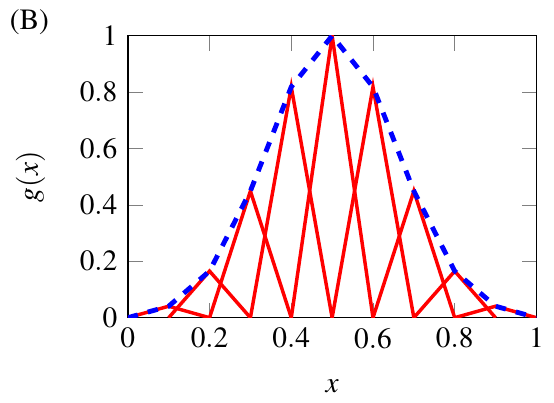}
\end{center}
\caption{\label{fig:FEM} Approximation by linear finite elements. (A) Linear basis functions \(\varphi_i\) 
on the unit interval. (B) Finite element approximation of the Gaussian \(g(x)\).}
\end{figure}

A suitable basis in one spatial dimension is, for instance, 
given by the linear hat functions, see Figure~\ref{fig:FEM}.A for a visualisation. 
Figure~\ref{fig:FEM}.B shows how the function \(g(x)\isdef\exp\big(-20(x-0.5)^2\big)\) 
is represented in this basis on the grid \(x_i=i/10\) for \(i=1,\ldots,9\). For the particular choice
of the hat functions, the coefficients are given by the evaluations of \(g\) at the nodes \(x_i\), i.e.\
\[
g(x)\approx\sum_{i=1}^9g_i\varphi_i(x)\quad\text{with }g_i\isdef\exp\big(-20(x_i-0.5)^2\big).
\]
Note that using the hat functions as a basis results in a tridiagonal matrix \({\bf A}\).

The presented approach can be transferred one-to-one to two and three 
spatial dimensions and also to more complex (partial) differential equations.
In practice, the \textit{ansatz space} \(V_N\) is obtained by introducing a 
triangular mesh for the domain in two spatial dimensions or a tetrahedral mesh
in three spatial dimensions, and then considering piecewise polynomial 
functions with respect to this mesh. To reduce the computational effort,
the basis functions are usually locally supported, since this results in a 
sparse pattern for the matrix \({\bf A}\).
%%%%%%%%%%%%%%%%%%%%%%%%%%%%%%%%%%%%%%%%%%%%%%%%%%%%%%%%%%%%%%%%%%%%%%%%%%%%%%%%%%%%%%%%%
%\input{ALE}
%!TEX root = ReviewArticle.tex
\tikzset{pics/.cd,
grid/.style args={(#1)#2(#3)#4(#5)#6(#7)#8}{code={%
\tikzset{pics/grid/dimensions=#8}%
% still don't get what the template does, so I fixed coordinates
% the corners of the grid are now a b c d
\coordinate (a) at (#1);
\coordinate (b) at (#3);
\coordinate (c) at (#5);
\coordinate (d) at (#7);
\foreach \i in {0,...,\y}
  \draw [pic actions/.try] ($(a)!\i/\y!(d)$) -- ($(b)!\i/\y!(c)$);
\foreach \i in {0,...,\x}
  \draw [pic actions/.try] ($(a)!\i/\x!(b)$) -- ($(d)!\i/\x!(c)$);

\path (#1) coordinate (-1) (#3) coordinate (-2)
      (#5) coordinate (-3) (#7) coordinate (-4);
}},
grid/dimensions/.code args={#1x#2}{\def\x{#1}\def\y{#2}}
}
\tikzset{pics/.cd,
prtcl/.style args={(#1)#2(#3)#4(#5)#6(#7)#8}{code={%
\tikzset{pics/prtcl/dimensions=#8}%
% still don't get what the template does, so I fixed coordinates
% the corners of the grid are now a b c d
\coordinate (a) at (#1);
\coordinate (b) at (#3);
\coordinate (c) at (#5);
\coordinate (d) at (#7);

\foreach \i in {1,...,\x} {
	\foreach \j in {1,...,\y} {
	\pgfmathsetmacro\weightx{(\i-0.5)/\x}
	\pgfmathsetmacro\weightxx{1-\weightx}
	\pgfmathsetmacro\weighty{(\j-0.5)/\y}
	\pgfmathsetmacro\weightyy{1-\weighty}
		\fill[pic actions/.try] ($\weighty * \weightx*(a)+\weighty*\weightxx*(b) 
		+ \weightyy * \weightx*(d)+\weightyy *\weightxx*(c)$) circle [radius=2pt];
		}
		}
}},
prtcl/dimensions/.code args={#1x#2}{\def\x{#1}\def\y{#2}}
}
\subsection{Arbitrary Lagrangian-Eulerian (ALE)}
The underlying idea of the \textit{Arbitrary Lagrangian-Eulerian} 
(ALE) description of motion is to decouple the movement of a given body
\(\Omega_0\subset\mathbb{R}^d\) from the motion of the underlying mesh 
that is used for the numerical discretisation. We refer to 
\cite{Michael:DHPR99} and the references therein for a comprehensive introduction into ALE. 
As motivated by the paragraph on continuum mechanics, 
we start from a deformation field \[\boldsymbol{\chi}\colon\Omega_0\times[0,\infty)\to\mathbb{R}^d,\]
which describes how the body \(\Omega_t=\boldsymbol{\chi}(\Omega_0,t)\) evolves and moves over time.
Remember that the position of a particle \({\bf X}\in\Omega_0\) at time \(t\geq0\) 
in Eulerian coordinates is given by \({\bf x}=\boldsymbol{\chi}({\bf X},t)\), whereas
its Lagrangian coordinates read \({\bf X}=\boldsymbol{\chi}^{-1}({\bf x},t)\).
Analogously, the corresponding \textit{velocity fields} for \(\Omega_t\) are then given by
\[
{\bf V}({\bf X},t)=\frac{\partial}{\partial t}\boldsymbol{\chi}({\bf X},t)
\]
in Lagrangian coordinates and by
\[
{\bf v}({\bf x},t)\isdef{\bf V}\big(\boldsymbol{\chi}^{-1}({\bf X},t),t\big)
\]
in spatial coordinates, respectively.

Usually, the representation of quantities of interest changes with 
their description in either spatial or material coordinates. 
Consider the scalar field \(u\colon\Omega_t\to\mathbb{R}\). We set
\[
\dot{u}({\bf x},t)\isdef\frac{\partial}{\partial t}u\big(\boldsymbol{\chi}({\bf X},t),t\big)\bigg|_{{\bf X}
=\boldsymbol{\chi}^{-1}({\bf x},t)},
\]
i.e.\ \(\dot{u}({\bf x},t)\) is the time derivative of \(u\) 
where we keep the material point \({\bf X}\) fixed. 
Therefore, \(\dot{u}({\bf x},t)\) is referred to as \textit{material derivative} of \(u\).
The chain rule of differentiation now yields
\begin{equation}
\dot{u}({\bf x},t)=\frac{\partial}{\partial t}u({\bf x},t)+{\bf v}({\bf x},t)\cdot\nabla u({\bf x},t).
\end{equation}
Thus, the material derivative \(\dot{u}({\bf x},t)\) 
is comprised of the \textit{spatial derivative} \(\frac{\partial}{\partial t}u({\bf x},t)\) 
and the advection term \({\bf v}({\bf x},t)\cdot\nabla u({\bf x},t)\),
see e.g.\ \cite{Michael:Hol00} for further details.
 
\begin{figure}[htb]
\begin{center}
\includegraphics[scale=0.45]{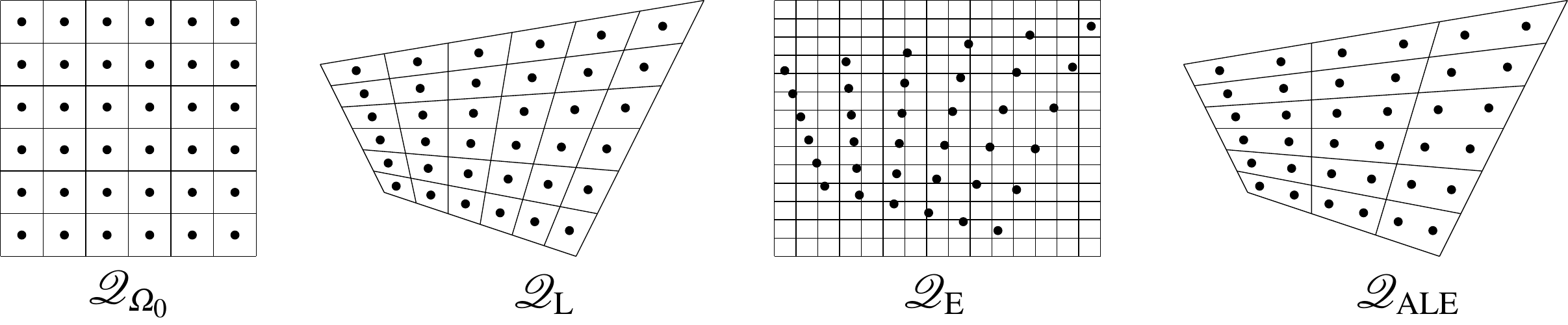}
\end{center}
\caption{\label{fig:ALEcoordinates}Comparison of the different frameworks:
Reference mesh with particles ($\mathcal{Q}_{\Omega_0}$), 
deformed mesh with particles in
the Lagrangian framework ($\mathcal{Q}_{\text{L}}$), 
fixed mesh with particles in the Eulerian framework ($\mathcal{Q}_{\text{E}}$),
modified mesh with particles in the ALE framework ($\mathcal{Q}_{\text{ALE}}$).}
\end{figure}
%\begin{figure}[htb]
%\begin{center}
%\scalebox{0.45}{\begin{tikzpicture}
%\pic at (0,0) [black] {grid={(0,0) (4,0)  (4,4) (0,4)  6x6}};
%\pic at (0,0) [black] {prtcl={(0,0) (4,0)  (4,4) (0,4)  6x6}};
%\pic at (6,1) [black]  {grid={(0,0) (3,-1) (5,3) (-1,2) 6x6}};
%\pic at (6,1) [black]  {prtcl={(0,0) (3,-1) (5,3) (-1,2) 6x6}};
%\pic at (12,0) [black] {grid={(0.1,0) (5.2,0)  (5.2,4) (0.1,4)  15x14}};
%\pic at (12.7,1) [black]  {prtcl={(0,0) (3,-1) (5,3) (-1,2) 6x6}};
%\pic at (19.5,1) [black]  {grid={(0,0) (3,-1) (5,3) (-1,2) 3x6}};
%\pic at (19.5,1) [black]  {prtcl={(0,0) (3,-1) (5,3) (-1,2) 6x6}};
%\draw (2,-0.6) node {\huge$\mathcal{Q}_{\Omega_0}$};
%\draw (8.5,-0.6) node {\huge$\mathcal{Q}_{\text{L}}$};
%\draw (14.6,-0.6) node {\huge$\mathcal{Q}_{\text{E}}$};
%\draw (22,-0.6) node {\huge$\mathcal{Q}_{\text{ALE}}$};
%\end{tikzpicture}}
%\end{center}
%\caption{\label{fig:ALEcoordinates}Comparison of the different frameworks:
%Reference mesh with particles ($\mathcal{Q}_{\Omega_0}$), 
%deformed mesh with particles in
%the Lagrangian framework ($\mathcal{Q}_{\text{L}}$), 
%fixed mesh with particles in the Eulerian framework ($\mathcal{Q}_{\text{E}}$),
%modified mesh with particles in the ALE framework ($\mathcal{Q}_{\text{ALE}}$).}
%\end{figure}

As a consequence, given the fixed computational mesh in the Eulerian framework, 
the domain \(\Omega_t\) moves over time. 
The Eulerian description is well suited to capture large distortions.
However, the resolution of interfaces and details becomes rather costly.
On the other hand, in the Lagrangian framework, it is easy to track free surfaces and interfaces, 
whereas it is difficult to handle large distortions, which usually require frequent remeshing of the domain
\(\Omega_0\). In order to bypass the drawbacks of both frameworks, the ALE method has been introduced, see
 \cite{Michael:DHPR99,Michael:HAC74}. Here, the movement of the mesh is decoupled from the movement 
 of the particles \({\bf X}\in\Omega_0\).  The mesh might, for example, be kept fixed 
 as in the Eulerian framework or be moved as in the Lagrangian framework or even be handled 
 in a completely different manner, see Figure~\ref{fig:ALEcoordinates} for a visualisation.

Within the ALE framework, the reaction diffusion equation on growing domains can be written as 
\[
\frac{\partial}{\partial t}c + {\bf w}\nabla c+ c\operatorname{div}{\bf v}=D\Delta c + R(c).
\]
Herein, \({\bf w}={\bf v}-{\bf u}\) the relative velocity between the material velocity 
\({\bf v}\) and the mesh velocity \({\bf u}\).
If the velocity of the mesh and the material coincide, i.e.\ \({\bf u} = {\bf v}\), 
the Lagrangian formulation is recovered. On the other
hand, setting \({\bf u}={\bf 0}\), we retrieve the Eulerian formulation, 
i.e.\ equation \eqref{eq:rdgrowth}, cf.\ \cite{Michael:ITFG+14,Michael:MMNI16}.
In practice, the mesh velocity is chosen such that one obtains a 
Lagrangian description in the vicinity of moving boundaries and an 
Eulerian description in static region, where a smooth transition 
between the corresponding velocities is desirable, cf.\ \cite{Michael:DHPR99}.
We remark that, in this view, the treatment of composite domains demands for additional care, 
particularly, when the subdomains move with different velocities, see \cite{KUMI14,Michael:MI12}.
%%%%%%%%%%%%%%%%%%%%%%%%%%%%%%%%%%%%%%%%%%%%%%%%%%%%%%%%%%%%%%%%%%%%%%%%%%%%%%%%%%%%%%%%%
%\input{LevelSet}
\subsection{Diffuse-Domain method}

\begin{figure}[htb]
\begin{center}
\includegraphics{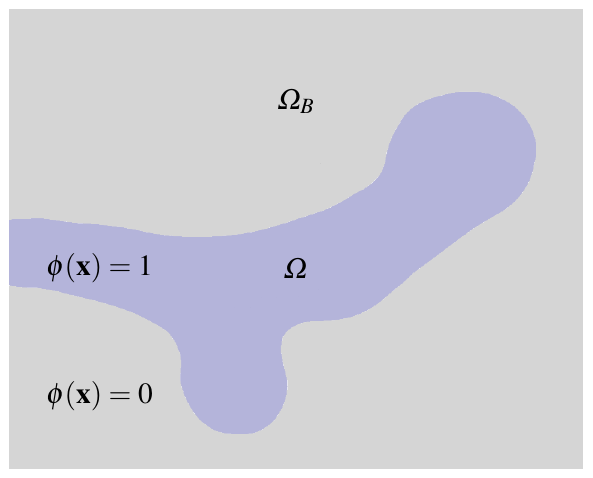}
\end{center}
\caption{\label{fig:phaseField} Visualisation of the computational domain \(\Omega\), 
the bounding box \(\Omega_B\) and the phase field \(\phi\).}
\end{figure}
%\begin{figure}[htb]
%\begin{center}
%\begin{tikzpicture}
%\draw (0,0) node {\includegraphics[scale=.3, clip, trim= 70 50  100 50]{./pics/DiffuseDomainImage}};
%\draw (0,1.4) node {\(\Omega_B\)};
%\draw (0,-0.3) node {\(\Omega\)};
%\draw (-2,-0.3) node {\(\phi({\bf x})=1\)};
%\draw (-2,-1.6) node {\(\phi({\bf x})=0\)};
%\end{tikzpicture}
%\end{center}
%\caption{\label{fig:phaseField} Visualisation of the computational domain \(\Omega\), 
%the bounding box \(\Omega_B\) and the phase field \(\phi\).}
%\end{figure}

The ALE method facilitates the modelling of moving and growing domains. 
Due to the underlying discretisation of the simulation domain, 
the possible deformation is still limited and topological changes cannot be handled. 
The \textit{Diffuse-Domain} method, introduced in \cite{Lucas:Kockelkoren2003-rf}, 
decouples the simulation domain from the underlying discretisation. 
A diffuse implicit interface-capturing method is used to represent the boundary 
instead of implicitly representing the domain boundaries by a mesh. 

The general idea is appealingly simple. We extend the integration domain to a 
larger computational bounding box and introduce an auxiliary field variable 
$\phi$ to represent the simulation domain. A level-set of $\phi$ describes 
the implicit surface of the domain, see Figure~\ref{fig:phaseField} for a visualisation. 
To restrict the partial differential equations to the bulk and/or surface, 
we multiply those equations in the weak form by the characteristic 
functions of the corresponding domain. To deform and grow the geometry, 
an additional equation with an advective term is solved to update the auxiliary field, 
cf.\ \cite{Lucas:Li2009-sa}. Following the example given in \cite{Lucas:Li2009-sa}, 
let us consider the classical Poisson equation with Neumann boundary conditions in the domain $\Omega$.
\begin{equation}\label{eq:poisson}
\begin{aligned}
     \Delta u &= f && \text{in} \ \Omega, \\
     \nabla u \cdot \vec{n} &= g && \text{on} \ \Gamma_\Omega.
\end{aligned}
\end{equation}
First, we extend the Poisson equation into the bounding box $\Omega_B\supset\Omega$. 
Thus, equation \eqref{eq:poisson} becomes
\begin{align}
    \operatorname{div}(\phi \nabla u) + \text{B.C.} = \phi f && \text{in} \ \Omega_.
\end{align}
Now, the Neumann boundary conditions can be enforced by replacing the $\text{B.C.}$-term by  
$\text{B.C.} = g |\nabla\phi|$  or $\text{B.C.} = \epsilon g |\nabla \phi|^2$, 
where $|\nabla\phi|$ and $\epsilon|\nabla\phi|^2$ approximate the Dirac $\delta$-function. 
Dirichlet and Robin boundary conditions can be dealt with similarly, see \cite{Lucas:Li2009-sa}. 
It is possible to show that in the limit $\epsilon \rightarrow 0$ the explicit 
formulation \eqref{eq:poisson} is recovered, cf.\ \cite{Lucas:Lervag2014-pg, Lucas:Li2009-sa}. 
Depending on the approximation of the $\delta$-function and the type of boundary conditions used,
the Diffuse-Domain method is first- or second-order accurate. 
The treatment of dynamics on the surface is more intricate as we do not have an explicit boundary anymore. 
The general idea is to extend the dependent variables constantly in surface-normal direction over the interface, 
cf.\ \cite{Lucas:Lowengrub2016-ou, Lucas:Wittwer2016-wj}.

Several methods exist to track the diffuse interface of the computation domain. 
In the level set method, the interface $\Gamma$ is represented by the zero isosurface 
of the signed-distance function to the surface $\Gamma_\Omega$.
In contrast to this artificial level set formulation, 
phase fields are constructed by a physical description of the free energy of the underlying system. 
Several such physical descriptions exist, but probably the most famous derivations are known under 
the Allen-Cahn and Cahn-Hilliard equations. 
The former formulation does not conserve the phase areas (or volumes in 3D) 
whereas the latter conserves the total concentration of the phases. 
Starting with a two phase system without mixing, the free energy can be described 
by the Ginzburg-Landau energy according to
\[ E_\sigma(\phi)\isdef\int_\Omega \frac{1}{\epsilon} W(\phi) + 
\frac{\epsilon}{2} | \nabla \phi|^2\operatorname{d}\!{\bf x}, 
\]
where $\Omega$ is the domain, $\epsilon$ controls the thickness of the interface, 
and $\phi$ is the phase field, cf.\ \cite{Lucas:Aland2012-xe}. 
Moreover, the function $W(\phi)$ is a double-well potential 
having its two minima in the values representing the bulk surfaces, e.g.\
\[ W(\phi) = \frac{1}{4} \phi^2(1-\phi)^2, \]
having its two minima at $\phi = 0$ and $\phi = 1$. 
The second term of the Ginzburg-Landau energy penalises gradients 
in the concentration field and thus can be interpreted 
as the free energy of the phase transition, see \cite{Lucas:Eck2011-cu}. 
With $W(\phi)$ defined as above, $E_\sigma$ is also referred 
to as surface energy or Cahn-Hilliard energy, cp.\ \cite{Lucas:Aland2012-xe}. 
Minimising the energy $E_\sigma$ with respect to $\phi$ results in solving the equation
\[ 
\frac{\operatorname{d}\!E_\sigma}{\operatorname{d}\!\phi}=\frac{1}{\epsilon} W'(\phi) 
- \epsilon \Delta \phi  \stackrel{!}{=} 0.
\]
The solution of this equation has two homogenous bulks describing the two phases 
and a $\tanh$-profile in between, see e.g.\ \cite{Lucas:Aland2012-xe}. 
The Allen-Cahn equation then reads as 
\begin{align*}
    \frac{\partial \phi}{\partial t} + {\bf v} \cdot \nabla \phi & = -M \mu(\phi),\\
    \mu(\phi) & = \frac{1}{\epsilon} W'(\phi) - \epsilon \Delta \phi,
\end{align*} 
where $M$ is a mobility parameter describing the stiffness of the interface 
dynamic and ${\bf u}$ the velocity field moving and deforming the phases again. 
An interface $\Gamma$ can be represented by the $\phi(\vec{x}) = 0.5$ contour 
line (iso-surface in 3D). Based on the interface profile, the surface normals can be approximated by
\[ \vec{n} = \frac{\nabla \phi}{|\nabla \phi|} \]
and the mean curvature by 
\[ \kappa = \nabla \cdot \vec{n}, \]
cf.\ \cite{Lucas:Aland2012-xe}.

The above description of the physical two-phase system minimises the area 
of the interface and thus exhibits unwanted self-dynamics for an interface tracking method. 
In\ \cite{Lucas:Folch1999-yn} it is proposed to add the correction term $-\epsilon^2\kappa(\phi)|\nabla \phi|$ 
to the right hand side, since $\kappa(\phi)$ depends on the local curvature, canceling out this effect, 
which is known as the Allen-Cahn law. Hence, the right-hand side becomes
\[ \mu(\phi) = \frac{1}{\epsilon} W'(\phi) - \epsilon \operatorname{div}(\vec{n}\vec{n} \cdot \nabla \phi).\]

The Diffuse-Domain method has been successfully applied to several biological problems. 
For example for modelling mechanically induced deformation of bones \cite{Lucas:Aland2014-ki} 
or simulating endocytosis \cite{Lucas:Lowengrub2016-ou}. 
It is even possible to couple surface and bulk reactions for modelling transport, 
diffusion and adsorption of any material quantity in \cite{Lucas:Teigen2009-yv}. 
Two-phase flows with soluble nanoparticles or soluble surfactants have been modelled 
with the Diffuse-Domain method in \cite{Lucas:Aland2011-ui} and \cite{Lucas:Teigen2011-me}, 
respectively. A reaction-advection-diffusion problem combining volume and surface diffusion 
and surface reaction has been solved with the Diffuse-Domain method in 
\cite{Lucas:Wittwer2016-wj, Lucas:Wittwer2017-ru} to address a patterning problem in murine lung development. 
%%%%%%%%%%%%%%%%%%%%%%%%%%%%%%%%%%%%%%%%%%%%%%%%%%%%%%%%%%%%%%%%%%%%%%%%%%%%%%%%%%%%%%%%%
%\input{LatticeBoltzmann}
\subsection{Lattice Boltzmann method}\label{ssec:LB}
The Lattice Boltzmann method (LBM) is a numerical scheme to simulate fluid dynamics.
It evolved from the field of cellular automata, more precisely lattice gas cellular automata 
(LGCA), in the 1980s.
The first LGCA that could simulate fluid flow was proposed in 1986, cp.\ \cite{Frisch1986}.
However, LGCA were facing several problems such as a relatively high fixed viscosity 
and an intrinsic stochastic noise, see \cite{frouzakis2011}.
In 1988, the LBM was introduced as an independent numerical method for fluid flow simulations, 
when tackling the noise problem of the LGCA method, cf.\ \cite{McNamara1988}.
A detailed description of LGCA and the development 
of the LBM from it can be found in \cite{Wolf-Gladrow2000}.

The LBM is applied in many different areas that study various different types 
of fluid dynamics for example incompressible, 
isothermal, non-isothermal, single- and multi-phase flows, etc., 
as well as biological flows, see \cite{frouzakis2011}.
As we have seen in the previous sections, biological fluid dynamics can be the
dynamics of biomolecules, i.e.\ morphogens, being solved in a fluid or the dynamics of
a biological tissue which can be approximated by a viscous fluid, cp.\ \cite{Michael:FFSS98}.
Further, the LBM can be used to simulate signaling dynamics of biomolecules 
by solving reaction-diffusion equations on the lattice.

In contrast to many conventional approaches modeling fluid flow on the macroscopic scale, 
for example by solving the Navier-Stokes equations, the LBM is a mesoscopic approach.
It is also advantageous for parallel computations because of its local dynamics.
LBM models the fluid as fictive particles that propagate and collide on a discrete lattice, 
where the incompressible Navier-Stokes equations can be captured 
in the nearly incompressible limit of the LBM, see \cite{Chen1998}.

The Boltzmann equation originates from statistical physics and describes 
the temporal evolution of a probability density distribution function
 $f({\bf{x}},{\bf{v}},t)$ defining the probability of finding a particle with
  velocity ${\bf{v}}$ at location ${\bf{x}}$ at time \(t\).
In the presence of an external force ${\bf{F}}$ acting on the particles and considering two processes, 
i.e.\ propagation of particles and their collision, the temporal evolution of $f({\bf{x}},{\bf{v}},t)$ is defined by 
\begin{equation*}
\frac{\partial f}{\partial t} + {\bf{v}}\cdot \nabla_{{\bf{x}}} f + {\bf{F}}\cdot \nabla_{{\bf{v}}} f = Q(f).
\end{equation*}
The most commonly used collision term is the single-relaxation-time Bhatnagar-Gross-Krook (BGK) collision operator
\begin{equation*}
Q(f)=Q_{\operatorname{BGK}}\isdef - \frac{1}{\tau} (f-f^{\operatorname{eq}}),
\end{equation*}
with $f^{\operatorname{eq}}$ being the Maxwell-Boltzmann equilibrium distribution 
function with a characteristic time scale $\tau$.
The discretized Boltzmann equation is then
\begin{equation*}
\frac{\partial f_i}{\partial t} + {\bf{v}}_{i,\alpha} \frac{\partial f_i}{\partial x_{\alpha}} 
= - \frac{1}{\tau} (f_i-f_i^{\operatorname{eq}}) + {\bf{F}}_i,
\end{equation*}
where $i$ determines the number of discrete velocities and  $\alpha$ the spatial dimensionality of the system.
The discrete LB equation is given by
\begin{equation}\label{eq:LBM}
f_i({\bf{x}}+{\bf{v}}_i \Delta t, t+ \Delta t) -f_i({\bf{x}}, t) 
=  - \frac{1}{\tau}\big(f_i({\bf{x}}, t)-f_i^{\operatorname{eq}}({\bf{x}}, t)\big) + {\bf{F}}_i. 
\end{equation}
The equation implicates a two step algorithm for the LBM.
In the first step referring to the left hand side of \eqref{eq:LBM}, 
the probability density distribution functions perform a free flight to the next lattice point.
In the second step referring to the right hand side of \eqref{eq:LBM}, 
the collision of the incoming probability density distribution functions on each lattice point is computed, 
followed by the relaxation towards a local equilibrium distribution function.
As the collision step is calculated on the lattice points only, 
the computations of the LBM are local, rendering it well suited for parallelisation. 
For further notes on the computational cost of the LBM, see \cite{frouzakis2011}.

The equilibrium function is defined by a second-order expansion of the Maxwell equation in terms of low fluid velocity
\begin{equation*}
f^{\operatorname{eq}} = \rho w_i \left[ 1
+ \frac{{\bf{v}}_i \cdot {\bf{u}}}{c_s^2} 
+ \frac{({\bf{v}}_i \cdot {\bf{u}} )^2}{2c_s^4}
- \frac{{\bf{u}} \cdot {\bf{u}} }{c_s^2}
\right]
\end{equation*}
with the fluid velocity ${\bf{u}}$, the fluid density $\rho$, the speed of sound $c_s$ and 
the weights $w_i$ that are given according to the chosen lattice, cf.\ \cite{He1997}.
The lattices used in LBM are regular and characterized as $DdQq$, 
where $d$ indicates the spatial dimension and $q$ the number of discrete velocities.
For simulations, commonly used lattices are $D1Q3$, $D2Q9$ and $D3Q19$, see \cite{frouzakis2011}.
The macroscopic quantities, i.e.\ density $\rho$ and momentum density $\rho {\bf{u}}$, 
are defined by the first few moments of the probability density distribution function $f_i$, i.e.\
\begin{equation*}
\rho = \sum_{i=0}^{q} f_i, \qquad \rho {\bf{u}} = \sum_{i=0}^{q} f_i {\bf{v}}_i.
\end{equation*}
The fluid pressure $p$ is related to the mass density $\rho$ via the equation for an ideal gas $p = \rho c_s^2$.
To simulate fluid-structure interactions, the LBM can be combined with the Immersed Boundary method (IBM), 
see \cite{Peskin2002}, that represents elastic structures being immersed in a fluid.
A detailed description of the IBM can be found in \cite{Tanaka2015}.
The LBM and IBM were first combined in \cite{Feng2004}, and later used, for instance, 
to simulate red blood cells in flow, cf.\ \cite{Zhang2007}, and to simulate the coupled tissue 
and signaling dynamics in morphogenetic processes with cellular resolution, cf.\ \cite{Tanaka2015b}.

%%%%%%%%%%%%%%%%%%%%%%%%%%%%%%%%%%%%%%%%%%%%%%%%%%%%%%%%%%%%%%%%%%%%%%%%%%%%%%%%%%%%%%%%%
%\input{Conclusion}
\section{Conclusion}
In this chapter, we have given an overview of the mathematical modelling of tissue dynamics, 
growth, and mechanics in the context of morphogenesis.
These different aspects of morphogenesis can be modeled either on a microscopic scale, 
for example by agent based models, or on a macroscopic scale
by continuum approaches. In addition, we have discussed several numerical 
approaches to solve these models.
Due to the complex nature and coupling of the different aspects that are 
required to obtain realistic models, 
the numerical solution of these models is computationally expensive.
When it comes to incorporating measurement data, these large computational 
efforts can place a severe limitation.
Algorithms for parameter estimation are usually based on gradient descent 
or sampling and therefore require frequent solutions of the model.
As a consequence, incorporating measurement data may quickly become 
infeasible and efficient algorithms have to be devised.
%%%%%%%%%%%%%%%%%%%%%%%%%%%%%%%%%%%%%%%%%%%%%%%%%%%%%%%%%%%%%%%%%%%%%%%%%%%%%%%%%%%%%%%%%
\bibliographystyle{splncs}
\bibliography{literature}

\end{document}